\documentclass[final,3p,times,twocolumn]{elsarticle}
\usepackage{amssymb}
\usepackage{amsmath}
\usepackage{graphicx} 
\usepackage{bm}
\usepackage{color}
\usepackage[english]{babel}
\journal{Physica A}
\begin{document}
\begin{frontmatter}
\title{	Mobility of nanometer-size solutes in water driven by electric field }

\author[hd]{Mohammadhasan Dinpajooh}
\author[dvm]{Dmitry V.\ Matyushov\corref{cor1}}      
\ead{dmitrym@asu.edu}   
\cortext[cor1]{Corresponding author}
\address[hd]{School of Molecular Sciences, 
Arizona State University, PO Box 871604, Tempe, Arizona 85287}
\address[dvm]{Department of Physics and School of Molecular Sciences, 
         Arizona State University, PO Box 871504, Tempe, Arizona 85287}
\begin{abstract}
We investigate the mobility of nanometer-size solutes in water in a uniform external electric field. General arguments are presented to show that a closed surface cutting a volume from a polar liquid will carry an effective non-zero surface charge density when preferential orientation of dipoles exists in the interface. This effective charge will experience a non-vanishing drag in an external electric field even in the absence of free charge carriers. Numerical simulations of model solutes are used to estimate the magnitude of the surface charge density. We find it to be comparable to the  values typically reported from the mobility measurements. Hydrated ions can potentially carry a significant excess of the effective charge due to over-polarization of the interface. As a result, the electrokinetic charge can significantly deviate from the physical charge of free charge carriers. We propose to test the model by manipulating the polarizability of hydrated semiconductor nanoparticles with light. The inversion of the mobility direction can be achieved by photoexcitation, which increases the nanoparticle polarizability and leads to an inversion of the dipolar orientations of water molecules in the interface.           
\end{abstract}

\end{frontmatter}

\section{Introduction}
\label{sec:1}
Electrophoretic mobility is the drag experienced by a dissolved, usually colloidal, particle in a uniform external electric field. Mobility of oil drops and air bubbles in water has been known for a long time \cite{Wall:10} and is traditionally linked to preferential adsorption of ions. Their counterions form the diffuse double layer. The overall charge measured by mobility is determined by an incomplete compensation between the charge of the adsorbed ions and the part of the diffuse layer within the shear surface. The latter encircles the stagnant layer of the electrolyte moving together with the dissolved particle. While the overall force acting on the ions of the electrolyte is zero, the electrokinetic drag is the result of choosing a limited volume within the electrolyte, surrounding the colloidal particle, with an uncompensated charge. The dragging force is thus the product of the average charge $\langle Q_R\rangle$ within the shear surface with the electric field acting on the charges.  We show here that the idea of a limited volume cut from the liquid and producing an excess charge can be extended to the dipolar polarization of the interface. While the dielectric surrounding the nanoparticle is neutral overall, like the electrolyte in the standard models, the divergent polarization of the interface produces an uncompensated bound charge when integrated over a finite volume. 

The excess of the adsorbed charge over the diffuse-layer charge, i.e., uncompensated charge $\langle Q\rangle_R\ne 0$, is reflected in the sign of the $\zeta$-potential at the shear surface \cite{Hunter:81}.  A negative $\zeta$-potential, typically recorded for oil drops and air bubbles in water, has been attributed to the excess of the adsorbed negative charge, with the hydroxide anion being a long-time favorite \cite{Marinova:1996ud,Takahashi:2005kl,Beattie:09,Zimmermann:2010db}. 

Recent calculations \cite{Buch:2007kx,Baer:2012uq} and measurements by surface-sensitive second-harmonic generation techniques \cite{Petersen:2008ys,Vacha:2011ij,Yamaguchi:2015hd,Samson:2014fz} do not support excessive adsorption of hydroxide to the oil-water \cite{Vacha:2011ij} and air-water \cite{Petersen:2008ys,Yamaguchi:2015hd} interfaces. In addition, the total X-ray reflection fluorescence spectroscopy \cite{Shapovalov:2013kc} provides the upper estimate for the free surface charges at the air-water interface at the level of $0.002$ (e/nm$^{2}$). Depending on the pH and other conditions, this estimate is up to two orders of magnitude below the surface charge density of 0.02--0.4 (e/nm$^2$ ) extracted from mobility \cite{Marinova:1996ud,Beattie:09,Vacha:2011ij,Samson:2014fz}. It seems plausible that either the formalism of estimating the surface charge density from mobility requires modification or alternative mechanisms of mobility, not involving ion adsorption, might be involved. 

The possibility of charge-free electrophoretic mobility in water has been discussed in the literature \cite{Joseph:08,Knecht:10,BenAmotz:2011cx,Vacha:2011ij,Vacha:2012dm,Schoeler:2013gg,DMmp:14}. The main idea here is that the microscopic structure of the interface, allowing molecular order within the hydration layers, can either produce an effective electrokinetic charge, not related to charges of free carriers, or substantially modify the effect of adsorbed ions on the overall mobility. This proposal has faced two difficulties. From the theoretical side, there is no established framework of how to translate the microscopic structure of the interface, captured by atomistic numerical simulations, into the macroscopic current. Care is required in implementing correct cutoff/boundary conditions \cite{Bonthuis:2009kw,Bonthuis:10,SukAluru:10} and statistical ensembles adequately representing the conditions of mobility measurements (as discussed briefly below). In addition, the field strengths required to produce sufficient sampling in simulations are significantly higher than experimental fields \cite{Daub:09} and can potentially modify the structure of the solution. From the experimental side, it is not clear how to connect the results of surface-sensitive spectroscopies, which directly report on the polarization structure of the interface \cite{Tian:2014kt,Bonn:2015hr}, with measured mobilities. 

Here we address the calculation of the force acting on a nanometer-size particle dissolved in water and placed in a uniform external field. We do not directly calculate the current produced in response to the external field assuming that, once the force is known, the mobility can be calculated by applying standard equations of hydrodynamics \cite{Mazur:1951vx,Yamaguchi:2015hd} (as shown for the capillary flow in the Appendix). Mobility of the hydrated solute is typically expressed, through Smoluchovski's equation (Eq.\ \eqref{eq:2} below), in terms of either the $\zeta$-potential or its effective charge. We derive a relation between the effective mobility (electrokinetic) charge and the interfacial structure of the water dipoles represented by the first-order orientational order parameter of the interface. This parameter is in principle accessible by surface-sensitive spectroscopies \cite{Tian:2014kt,Bonn:2015hr,Strazdaite:2015hx,Wen:2016df} and by equilibrium computer simulations of solutions \cite{Giovambattista:2007kx,DMjcp2:11}.    

This model shows that the effective charge of the solute responding to the uniform external field is not equal to the charge of the free carriers. It is therefore possible that the effective electrokinetic charge reported by mobility measurements significantly overestimates the number of adsorbed ions. The orientational structure of interfacial dipoles is the key in understanding these differences. Since the interfacial structure and dipolar orientations in the interface can be altered by modifying the solute/substrate \cite{Giovambattista:2007kx,Nihonyanagi:2009vn,Vacha:2011ij}, one gains the means to experimentally test both the effect of the interface on the effective electrokinetic charge and the hypothesis of charge-free mobility. In particular, we suggest that changing the polarizability of a (semiconductor) nanoparticle by exciting electron-hole pairs can invert the sign of the mobility. This effect is driven by the relation between the orientation of dipoles in the hydration layer with the nanoparticle polarizability \cite{DMprl:15} manipulated by light \cite{WangHeinz:06}.     

\section{Interfacial structure and particle mobility}
\label{sec:2}

\subsection{General arguments}
We start by considering a single spherical ion with the charge $q$ at its center and with the radius $a$. It is placed in a polar liquid with the bulk static dielectric constant $\epsilon_s$. We will further consider a spherical liquid sample with the macroscopic radius $L$ and place the ion at its center to simplify the geometry. The instantaneous charge density in the sample is
\begin{equation}
\rho = \rho_i + \rho_b,
\label{eq:1}  
\end{equation}
where $\rho_i=q\delta(\mathbf{r})$ and $\rho_b(\mathbf{r}) = \sum_j q_j\delta \left(\mathbf{r}-\mathbf{r}_j \right)$ is the density of bound charge at a given instantaneous configuration of the liquid with the atomic partial charges $q_j$ located at the coordinates $\mathbf{r}_j$. Based on charge conservation \cite{Landau8}, $\rho_b=-\nabla \cdot \mathbf{P}$ is expressed in terms of the polarization density field $\mathbf{P}$. No specific approximation, such as the dielectric boundary value problem, is assumed here. The instantaneous polarization field is given by the microscopic expression \cite{Gubbins:84,Jackson:99} 
\begin{equation}
\mathbf{P}(\mathbf{r})= \sum_j  \mathbf{m}_j\delta\left(\mathbf{r}-\mathbf{r}_j\right) -\tfrac{1}{3} \nabla\cdot\sum_j \mathbf{Q}_j\delta\left(\mathbf{r}-\mathbf{r}_j\right) + \dots .
\label{eq:2}
\end{equation}
Here, $\mathbf{m}_j$ denotes the molecular dipole, $\mathbf{Q}_j$ is the molecular quadrupole (defined according to Ref.\ \cite{Gubbins:84}), and the dots refer to the higher-order multipolar terms. When the statistical average over the configurations of the liquid is performed, one arrives at statistically averaged scalar and vector fields, $\langle \rho_b\rangle$ and $\langle\mathbf{P}\rangle$.  

From Eq.\ \eqref{eq:1}, one can calculate the overall charge within a spherical volume $\Omega_R$ with the radius $R$ surrounding the ion at its center
\begin{equation}
\langle Q_R\rangle =\int_{\Omega_R} \left[\rho_i - \nabla\cdot\langle \mathbf{P}\rangle \right] d\mathbf{r} .
\label{eq:3}
\end{equation}
By using the Gauss theorem, integration in Eq.\ \eqref{eq:3} yields 
\begin{equation}
\langle Q_R\rangle = q + \left[S_a P_a - S_RP_R \right].
\label{eq:4}
\end{equation}
Here, $P_a$ and $P_R$ are the average radial projections of the polarization density, $P_r=\mathbf{\hat r}\cdot\langle\mathbf{P}\rangle$, $\mathbf{\hat r}=\mathbf{r}/r$ taken at $r=a$ and $r=R$, respectively. Further, $S_a=4\pi a^2$ and $S_R=4\pi R^2$ are the surface areas. The polarization is locally proportional to the field in the continuum electrostatics and $P_r= (q/4\pi r^2)(1-\epsilon^{-1})$. Therefore, in this case, $r^2P_r$ is independent of $r$ and the two summands in the brackets in Eq.\ \eqref{eq:4} cancel out. One gets $\langle Q_R\rangle=q$.  

The dielectric sample is overall neutral and one can additionally require
\begin{equation}
\int_\Omega \langle \rho_b\rangle d\mathbf{r} = 0,
\label{eq:5}  
\end{equation}
where the integral is taken over the liquid volume $\Omega$ between the spheres $r=a$ and $r=L$. This relation imposes the boundary condition
\begin{equation}
a^2P_a = L^2P_L,
\label{eq:6}  
\end{equation}
which is satisfied for continuum electrostatics.  

The normal, outward to the dielectric component of the polarization field $\sigma_q=P_n=-P_a$ plays the role of the surface charge density of a discontinuous dielectric interface \cite{Landau8}. While this charge originates from a divergent polarization of bound molecular charges, it is experimentally observable. To show that, one can consider the electrostatic potential created by free and bound charges inside or outside of the macroscopic sample. It is given as a sum of the electrostatic potentials arising from the free and bound charges \cite{EygesBook:72} 
\begin{equation}
\phi = \frac{q}{r} + \oint_{S_a} \frac{\sigma_q}{\left|\mathbf{r}-\mathbf{r}_S \right|} dS = \frac{q}{r} - \left(1-\frac{1}{\epsilon_s}\right)\frac{q}{r},
\label{eq:7}  
\end{equation}
where the surface integral is over the surface of the ion $S_a$. The overall potential $\phi=q/(r\epsilon_s)$ is said to be dielectrically screened. This physically implies that any probe charge placed outside of the dielectric sample will sense the combined charge $q_\text{eff}=q/\epsilon_s$, resulting from adding the ion charge with the opposite bound charges non-uniformly distributed around the ion and producing a non-zero divergence $\nabla\cdot \mathbf{P}$.     

\begin{figure}
\includegraphics*[clip=true,trim= 0cm 1.5cm 0cm 0cm,width=7cm]{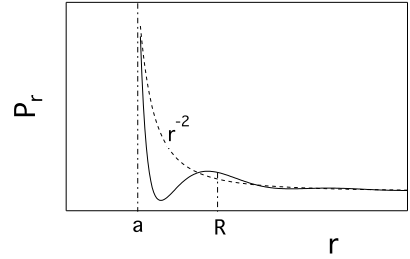}
\caption{Cartoon depicting the radial projection of the microscopic polarization density $P_r$ (solid line) and its dielectric form $\propto r^{-2}$ (dashed line). The volume integral of $\partial P_r/\partial r$ between surfaces $r=a$ and $r=R$ in Eq.\ \eqref{eq:3} can be non-zero, while it always vanishes in the dielectric limit. 
} 
\label{fig:1}  
\end{figure}

We now move to the next step to point out that the polarization field of liquid interfaces often shows a behavior more complex than $P_r \propto r^{-2}$ of continuum electrostatics \cite{Ballenegger:05,Horvath:2013fe,Bonthuis:2013fk}. The function $P_r$ often displays overscreening, which means that it can be much larger in the magnitude at the contact with the ion than predicted by dielectric models. It also shows oscillations caused by molecular granularity as it decays to the $r^{-2}$ asymptote at $r\to\infty$. While the overall neutrality condition \eqref{eq:6} still must hold, the charge $\langle Q\rangle_R$ obtained by integrating in Eq.\ \eqref{eq:3} over a small volume $\Omega_R$ can be nonzero for a function $P_r(r) = p(r)/r^2$ with a generally oscillatory $p(r)$ such that $p(\infty)=1$ (Fig.\ \ref{fig:1}). 

This simple observation is the basis of our proposed alteration of the standard model of ionic mobility under the drag of a uniform electric field. We suggest that $\langle Q\rangle_R\ne q$ if the liquid within the shear surface, dragged along with the ion, carries some molecular interfacial structure affecting the radial distribution of the polarization density. The effective charge associated with mobility is affected by the distribution of the bound charge within the shear surface, in addition to the total charge of free carriers.

\subsection{Ionic mobility}
\begin{figure}
\includegraphics*[clip=true,trim= 0cm 0cm 0cm 0cm,width=5.1cm]{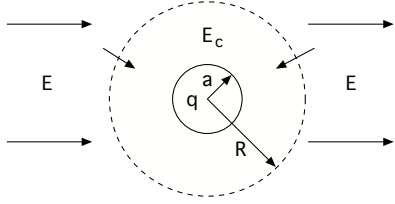}
\caption{Ion with the charge $q$ and the radius $a$ immersed in a polar liquid in the uniform macroscopic (Maxwell) field $E$. $R$ indicates the radius of the shear sphere incorporating the stagnant layer of the liquid dragged by the field along with the solute. $E_c$ is the cavity field of the uniformly polarized liquid created inside the shear sphere. The arrows on the opposite sides of the spherical cavity indicate water dipoles oriented  favorably (left) and unfavorably (right) along the external field. The difference in the chemical potential between right and left is positive. It creates an osmotic pressure pushing the particle in the  direction opposite to the field and corresponding to an effective negative charge.    
}
\label{fig:2}
\vskip -0.5cm
\end{figure}

The hydrodynamic mobility of an ion is determined by the shear surface of the radius $R$, which is coarse-grained to smooth out the details of molecular granularity by averaging out the molecular motions on the time short compared to the time-scale of hydrodynamic flow (Fig.\ \ref{fig:2}). Electrostatics suggests that the force acting on the ion and its stagnant layer is the product of the average charge $\langle Q_R\rangle$ within the shear surface with the field acting on these charges. This field is the cavity field $E_c$ \cite{Jackson:99} combining the field from external charges with the field of the polarized dielectric outside of the shear surface 
\begin{equation}
\langle F\rangle = \langle Q_R\rangle E_c .
\label{eq:8}
\end{equation}
In dielectric theories, the cavity field inside a sphere is related to the macroscopic Maxwell field $E$ by the equation \cite{Jackson:99} 
\begin{equation}
E_c=\frac{3\epsilon_s}{2\epsilon_s+\epsilon_p}E,
\label{eq:9}
\end{equation}
where $\epsilon_p$ is the dielectric constant of the particle.  

The steady flow of dissolved particles with the speed $u$ is reached when the electrostatic drag is counterbalanced by hydrodynamic friction, $\langle F\rangle = 6\pi\eta u R$, where $\eta$ is the bulk viscosity. The resulting mobility $\mu=u/E\simeq\langle Q_R\rangle /(4\pi\eta R)$ [$\epsilon_s\gg\epsilon_p$, $\epsilon_s\gg 1$ in Eq.\ \eqref{eq:1}] gives direct access to the total charge $\langle Q_R\rangle$. Smoluchovski's equation, typically used in the literature, re-writes this relation in terms of the $\zeta$-potential defined as the electrostatic potential at the shear surface $\zeta=\langle Q_R\rangle/(\epsilon_sR)$ \cite{Ohshima:06}. The result is the equation for the mobility \cite{Hunter:81}
\begin{equation}
\mu = \frac{\epsilon_s\zeta}{4\pi\eta} . 
\label{eq:9}
\end{equation}

This formalism is well established, and the results of mobility measurements are often cast in terms of the effective surface charge density $\sigma_\text{eff} = \langle Q_R\rangle/S$, where $S$ is the surface area of the particle. We follow this established practice and focus mostly on $\langle Q_R\rangle$ and the corresponding  $\sigma_\text{eff}$. The arguments given here need to be modified with the account for the diffuse potential when  electrolyte is present \cite{Ohshima:06}. We do not consider these effects here and focus instead on charged or uncharged particles dissolved in a polar molecular solvent, which establishes a microscopic multipolar structure in the interface. The main outcome of this perspective is the modification of the effective charge $\langle Q_R\rangle$ by the dipolar order of the interface expressed in terms of the average cosine of the interfacial dipoles (order parameter) $p_1$.

Starting from Eq.\ \eqref{eq:4}, one can proceed in two steps and first apply a reasonable approximation to the surface charge density at the shear surface. The surface charge density at the actual physical surface of the solute then becomes our main focus. Since the shear surface does not involve any physical disruption of the liquid structure, it is reasonable to assume that $P_R$ can be related to the field of the ion charge by the rules of continuum electrostatics \cite{Jackson:99} $S_RP_R = q(\epsilon_s-1)/\epsilon_s$. We stress that this assumption does not affect the main points of our reasoning, as will be clear from the discussion below.  With the continuum polarization at the shear surface one gets in Eq.\ \eqref{eq:5}  
\begin{equation}
\langle Q_R\rangle = q\epsilon_s^{-1} - \sigma_q S_a .
\label{eq:10}
\end{equation}
Since the microscopic susceptibility of the nanometer interface can significantly deviate from the rules of macroscopic continuum electrostatics \cite{Ballenegger:05,DMcpl:11,Bonthuis:2013fk}, $\sigma_q=P_n=-P_a$ is left unspecified in Eq.\ \eqref{eq:10}. The simple message delivered by Eqs.\ \eqref{eq:9} and \eqref{eq:10} is that asymmetry in the water susceptibility between the shear and solute dividing surfaces leads to a modification of the standard result $\langle Q_R\rangle=q$.   

Since $\sigma_q$ is given by the normal projection of the polarization density in the interface, Eq.\ \eqref{eq:10} offers a new result typically absent in standard dielectric models. Those suggest that $\sigma_q$  is proportional to the ion charge $q$. However, if the polar liquid is spontaneously polarized in the interface, i.e., if the interfacial dipoles possess preferential non-random orientations caused by the interfacial order \cite{Horvath:2013fe,Remsing:2014fo,Beck:2013gp}, $\langle Q_R\rangle\ne 0$ even at $q=0$. What is required is a nonzero radial projection of the dipolar polarization density at the solute surface.

The dipole ordering in the interface can be described by the first-order orientational order parameter  $p_1=\langle \mathbf{\hat m}\cdot \mathbf{\hat r}\rangle_a$, which is the average cosine of the water dipole moment projected on the radial direction and calculated in a narrow layer at the solute surface $r=a$ \cite{DMcpl:11,DMjcp2:11}. The surface charge density can be written in terms of the water dipole moment $m$ and the order parameter, $-\sigma_q = (m p_1/S) (d N_s/dr)\big|_{r=a}$. Here, $N_s=N_s(r)$ is the number of water molecules within the shell of the radius $r>a$.  By using the definition of the number of water molecules in the shell in terms of the solute-solvent radial distribution function (RDF) $g_{0s}(r)$, one can re-write $\sigma_q$ as
\begin{equation}
-\sigma_q = \rho m p_1 G,
\label{eq:11}
\end{equation}
where $G=g_{0s}(a)$ is the contact value of the solute-solvent RDF and $\rho$ is the number density of bulk water. Equation \eqref{eq:11} is written for an arbitrary value of $q$, which means that $p_1G$ should be calculated in the presence of the ion charge $q$; $\sigma_0$ corresponds to $q=0$. 

The value of $\sigma_q$ for large particles can be estimated from the $a\to \infty$ asymptote for the hard-sphere (HS) solute \cite{Luzar:1987gf} $G_\text{HS}\to \beta P/\rho$, which results in $-\sigma_q\to  \beta mp_1 P$, where $P$ is the hydrostatic pressure. This gives for water at ambient conditions $-\sigma_q\simeq  10^{-3} p_1(G/G_\text{HS})(P/\mathrm{atm})$ e/nm$^2$, where $G/G_\text{HS}\simeq \exp[-\beta\Delta\mu_w]$ defines the affinity of water toward the solute $\Delta\mu_w$ beyond simple HS packing preferences.   

Equation \eqref{eq:11} establishes the effective charge of a closed spherical interface within a polar liquid. Its sign is fully defined by the orientational order parameter $p_1$: it is negative when the water dipoles preferentially orient toward the solute/cavity and is positive when they point toward the bulk. This equation shows that any closed dividing surface, cutting a volume from a polar liquid, will be dragged by an external electric field if a preferential orientation of dipoles in the interface is established. This result is independent of the presence of the electrolyte since bound charges are not screened by the ions. 

The proposed formalism equally applies to the problem of a water drop in a nonpolar solvent (oil) \cite{Schoeler:2013gg}. Equation \eqref{eq:11} still defines the surface charge density with the convention that the orientational order parameter is calculated by projecting the surface water dipoles on the radial direction pointing toward water (inward in the case of a drop). To make our assignment clear, the water-oil interface with water's hydrogen pointing toward the oil phase \cite{Scatena:2001ve,Vacha:2011ij,Strazdaite:2014jc} will, according to Eqs.\ \eqref{eq:10} and \eqref{eq:11}, produce a negative charge of the water drop.

It is important to note that the electrostatic force linear in the external field, $\langle F\rangle \propto E_c$, assumes an unperturbed orientational structure of the interface projected on the order parameter $p_1$ \cite{com:Bonthuis}. The relaxation of the interfacial order in response to the external field would represent the interfacial polarizability, which contributes to the overall force as a term quadratic in the external field. We do not consider the interface polarizability here assuming that macroscopic fields used in experiment are weak compared to microscopic interfacial fields \cite{Horvath:2013fe} and do not significantly alter dipolar orientations in the interface.

The electrolyte is overall neutral and the overall force acting on the electrolyte ions is zero $F=\sum_i q_i E=0$. However, producing current requires work of the external source. The power $P$, or the rate of doing work, is related to the current density $\mathbf{j}$ \cite{Jackson:99} 
\begin{equation}
P=\int \mathbf{j}\cdot \mathbf{E}\, d\mathbf{r}= P_\text{el} + N_0\langle Q_R\rangle u E ,   
\label{eq:12}
\end{equation}
where $P_\text{el}=(J_+-J_-)E$ is the power required to move the electrolyte ions with the overall current of cations and anions given as $J_{\pm}$;  $N_0$ is the number of colloidal particles (see Supplemental Material for detail). Equations \eqref{eq:5} and \eqref{eq:12} in principle allow a non-zero current and power production at $q=0$, i.e., for overall neutral solutes surrounded by a polarized interface. This possibility was viewed in Ref.\ \cite{Bonthuis:11} as contradicting to Saxen relations between the streaming potential and electro-osmotic current, which are specific forms of the Onsager reciprocal relations \cite{Mazur:1951vx}. We show in the Appendix that the Onsager relations are obeyed in our model by the simple fact of being based on the Coulomb law applied to both free and bound charges.   

The drag experienced by a closed surface can be viewed as a specific form of osmosis \cite{Einstein05}. The gradient of the chemical potential of interfacial waters at the opposite sides of the surface is created by the external field. It is the consequence of the favorable orientation with the field of the molecules on one side of the surface compared to the unfavorable orientation on the opposite side \cite{Bratko:2007jl}  (surface arrows in Fig.\ \ref{fig:2}). The chemical potential gradient will result in the osmotic pressure difference on the opposite sides of the surface as long as spontaneous order in the interface persists. This physical interpretation of non-zero mobility implies that direct numerical simulations of this effect will require the $\mu$VT ensemble \cite{Luzar:1987gf,Bratko:2007jl}, keeping the chemical potential of water constant. Since these results are presently not available, we use more conventional NVT and NPT simulations of nonpolar and ionic solutes in water to estimate the interfacial charge density $\sigma_q$ in Eq.\ \eqref{eq:11} from the computed $p_1G$ parameter.

\begin{figure}
\includegraphics*[clip=true,trim= 0cm 1.8cm 0cm 0cm,width=6.1cm]{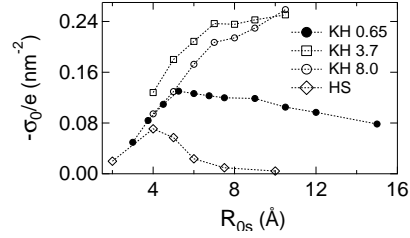}
\caption{Surface charge density of hard-sphere (HS) and Kihara (KH) solutes of varying solute size $R_{0s}$ in SPC/E water \cite{DMcpl:11} (solid points) and TIP3P water (open points). The LJ energy $\epsilon_{0s}$ in Eq.\ \eqref{eq:9} was varied in the simulations: 0.65, 3.7, and 8.0 kJ/mol. The dotted lines connect the points to guide the eye.    
}
\label{fig:3}
\vskip -0.3cm
\end{figure}

\section{Computer simulations}
\label{sec:3}
We have considered HS and Kihara (KH) solutes dissolved in force-field water. The Kihara potential is the HS core modified with the Lennard-Jones (LJ) layer at its surface \cite{Kihara:58}. Specifically, the solute-solvent potential is given as 
\begin{equation}
\phi_{0s}(r) = 4\epsilon_{0s}\left[\left( \frac{\sigma_{0s}}{r-R_\text{HS}}\right)^{12} - \left( \frac{\sigma_{0s}}{r-R_\text{HS}}\right)^{6}\right], 
\label{eq:13}
\end{equation}
where $\epsilon_{0s}$ in the LJ energy and $\sigma_{0s}$ is the distance between the solute HS core with the radius $R_\text{HS}$ and water's oxygen. $\sigma_{0s}=3$ \AA\ was kept constant in the simulations, while $R_\text{HS}$ and $\epsilon_{0s}$ were varied.  

The molecular dynamics (MD) and Monte Carlo (MC) simulations presented here address the question of whether the product $p_1G$ characterizing water interfacing these solutes will produce $\sigma_q$ comparable to experiment. The experimental $\sigma_q$ reported in the literature are derived from mobility through Smoluchovski's equation \cite{Beattie:09,Vacha:2011ij} [Eq.\ \eqref{eq:2}]. The details of the simulation protocols have been discussed elsewhere \cite{DMcpl:11,DMprl:15,DMjcp2:15} and are given in the Supplemental Material. Here we focus only on the results.  
 
Figure \ref{fig:3} shows  $\sigma_0/e$ ($q=0$, $e$ is the elementary charge) from the simulation data changing with the size of the HS and KH solutes in TIP3P and SPC/E water models \cite{tip3p:83}. The size of the KH solute is measured as $R_{0s}=R_\text{HS}+\sigma_{0s}$ [Eq.\ \eqref{eq:9}]. It approximates well the position of the first peak of the solute-solvent RDF. The size of the HS solute $R_{0s}$ is defined as the distance of the closest approach of the water oxygen to the solute. It gives the exact position of the RDF's first peak. 

The sign of the surface charge density is negative for both HS and KH solutes, reflecting the preferential orientation of the surface water dipoles into the bulk. Increasing the solute-solvent LJ attraction makes the hydration shell denser, as reflected by a higher $-\sigma_0$. The fast drop of $-\sigma_0$ for the HS solute is caused by its partial dewetting \cite{Sarupria:09} when $R_{0s}\geq 5$ \AA. 

The magnitude of $\sigma_0$ is somewhat higher than the values typically reported from mobility measurements ($\sim -0.04$ (e/nm$^{2}$) for hexadecane in 0.2 mM NaCl at $\mathrm{pH} = 7$ \cite{Beattie:09}). We estimated the $\zeta$-potential for the $\epsilon_{0s}=0.65$ kJ/mol Kihara solute \cite{DMcpl:11} (Fig.\ \ref{fig:3}). It turned out that $\sigma_0 R_{0s}^2$ is an approximately linear function of $R_{0s}$ (Fig.\ S3 in Supplemental Material) such that $\zeta \simeq 0.026 (\mathrm{e/nm}) (R_{0s}/R)$ for large Kihara solutes. Neglecting the difference between $R_{0s}$ and $R$ in this limit results in $\zeta\simeq 38$ mV. This number is not very different in magnitude from those typically reported experimentally. For instance, $\zeta\simeq - 81\pm14$ mV was reported for xylene droplets in 10$^{-5}$M NaCl electrolyte at $\mathrm{pH=6}$ \cite{Marinova:1996ud}. For water at room temperature, the Debye-H{\"u}ckel length is $\kappa^{-1}\simeq3/c^{1/2}$ \AA\ for a single-charge electrolyte with the molar concentration $c$ \cite{Falkenhagen}.  For the cited experiment, one gets $\kappa^{-1}\simeq 10^3$ \AA\ and the amount of counterion charge within the stagnant layer of $<1$ nm in thickness \cite{Knecht:2013jo} can be neglected. The measured $\zeta$-potential thus reflects the effective electrokinetic surface charge.  We stress that our solutes are significantly smaller in size than oil drops used in the mobility measurements ($\sim 100$ \cite{Samson:2014fz} to $\sim 200-300$ \cite{Vacha:2011ij} nm) and have a smooth surface, in contrast to the corrugated surface of oil emulsions \cite{Strazdaite:2014jc}.

The experimental $\zeta$-potential \cite{Marinova:1996ud} has the sign opposite to that calculated for the Kihara solutes. The reason is the positive sign of $p_1$ in the Kihara-water interface, while negative $p_1$ values have been recently reported for the oil-water interface 
\cite{Vacha:2011ij,Strazdaite:2015hx}. The access to water orientation in the interface is experimentally provided by heterodyne-detected vibrational sum-frequency generation (VSFG) spectroscopy through the imaginary part of the VSFG signal $\mathrm{Im}\chi^{(2)}$ \cite{Strazdaite:2015hx,Wen:2016df}. The combination of the sign  of $\mathrm{Im}\chi^{(2)}$ and its intensity in principle gives access to $p_1$, although in reality fitting of simulations to experimental spectra is required \cite{Pieniazek:2011fc}. Resolving all features of the experimentally reported  $\mathrm{Im}\chi^{(2)}$ requires including three-body interactions in the force field model of water \cite{Pieniazek:2011fc}. Whether the same is true regarding the values of $p_1$ is not clear at the moment, although there are indications that three-site models of water somewhat overestimate its spontaneous orientational structure in the interface \cite{Remsing:2014fo}.  In addition to spontaneous orientation in the uncharged interface, the orientation of water dipoles and corresponding $p_1$ are strongly affected by the presence of ions \cite{Wen:2016df,Roy:2014kl} as we discuss next. 

\begin{figure}
\includegraphics*[clip=true,trim= 0cm 1.5cm 0cm 0cm,width=6cm]{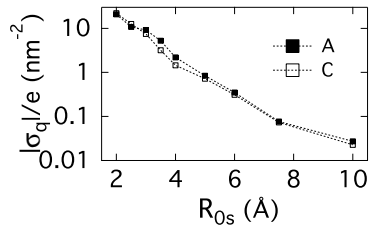}
\caption{Surface charge density of HS cations (C) and anions (A) in TIP3P water. The calculations are done according to Eq.\ \eqref{eq:11}; the dotted lines connect the points. $\sigma_q$ is negative for cations and positive for anions. }
\label{fig:4}
\vskip -0.5cm
\end{figure}

The potential situation with hydrated ions is illustrated in Fig.\ \ref{fig:4}, where $\sigma_q$ is calculated from Eq.\ \eqref{eq:11} for HS cations and anions of varying size in TIP3P water in the absence of counterions \cite{DMjcp2:15}. The main observation here is that $\sigma_q$ significantly exceeds in the absolute magnitude the prediction of the continuum electrostatics. This means $|\langle Q_R\rangle|\gg |q| $ in Eq.\ \eqref{eq:9}, which should lead to an overestimate of the number of adsorbed ions when the standard equations for the screening of free charge carriers in electrolytes are applied to the analysis of the mobility data \cite{Ohshima:06}.

The overpolarization of the water dipoles attached to the surface ions might have significant implications for the interpretation of the mobility data. Figure \ref{fig:4} indicates that the microscopic orientational order of the water dipoles next to a positive ion will significantly enhance its effective electrokinetic charge determined from the mobility measurement. Correspondingly, a negative adsorbed ion will appear more negative in the particle mobility. Therefore, the actual concentration of adsorbed ions can be significantly lower than estimated from mobility. This observation might help to explain the disagreement between the electrokinetic measurements \cite{Beattie:09,Vacha:2011ij,Samson:2014fz} and surface-sensitive spectroscopies \cite{Petersen:2008ys,Vacha:2011ij,Yamaguchi:2015hd,Samson:2014fz} regarding the concentration of the surface adsorbed ions.  The actual extent of overpolarization requires more extensive simulations in realistic electrolytes. One also should not underestimate the potential effect of corrugation of any real water-oil interface \cite{Strazdaite:2014jc}, which will affect the average contact RDF $G$ in Eq.\ \eqref{eq:11}. 

Figure \ref{fig:4} indicates that surface charge densities of large positive and negative ions with $|q|=1$ charge at the center are close in magnitude. However, this outcome might not hold for small ions adsorbed at the surface of a large particle. The product $p_1G$ is generally asymmetric between cations and anions because of the asymmetry in the charge distribution of the water molecule \cite{Lynden-Bell:1997uq}. In that case, the orientational order and the corresponding surface charge density will not compensate between the oppositely charged adsorbed ions, and a non-vanishing $\sigma_q$ will follow even at the total zero charge. The observable consequence of this asymmetry would be a shift between the iso-electric point of electrokinetic mobility and the point of zero charge, as reported for some systems \cite{Drzymala:1999bs}. Overall, the main result of the general formalism summarized by Eq.\ \eqref{eq:10} and simulations performed here is that the charge of free carriers and the effective electrokinetic charge incorporating the interfacial dipolar order can be significantly different.

\begin{figure}
\includegraphics*[clip=true,trim= 0cm 1.4cm 0cm 0cm,width=6cm]{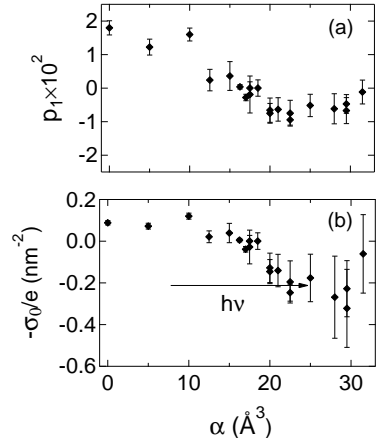}
\caption{The order parameter $p_1$ (a) and the surface charge density $\sigma_0$ (b) vs the polarizability of a polarizable HS solute, carrying the isotropic polarizability $\alpha$ and dipole moment $m_0=5$ D, in TIP3P water \cite{DMprl:15} (error bars show the uncertainties of calculations).  The horizontal arrow indicates the photoinduced alteration of the polarizability that inverts the mobility of the nanoparticle. }
\label{fig:5}
\vskip -0.5cm
\end{figure}

\section{Experimental testing and conclusions}
\label{sec:4}
In conclusion, we have derived a simple equation [Eqs.\ \eqref{eq:10} and \eqref{eq:11}] relating the effective charge of a hydrated nanoparticle to the orientational order in the interface and the water density in contact with the solute. Both parameters carry asymmetry between positive and negative charges. Therefore, the surface charge density $\sigma_q$ induced by the positive and negative free carriers will not compensate and produce an overall nonzero value even when the total charge is zero. The electrokinetic charge can be substantially enhanced by the dipolar order in the interface and the theory predicts a non-zero effective charge when the interface is spontaneously polarized in the absence of free charge carriers (charge-free mobility). The values of the surface charge density derived from simulations of uncharged nanometer-size solutes are consistent or exceed those typically reported from mobility measurements. 

Our development traces in spirit the well-established mechanism of electrophoretic mobility due to ions of electrolyte. Both ions and the dipoles of the solution surrounding the colloidal particle are neutral overall. However, there is an excess of ions within the shear surface of the particle, which results in the electrokinetic charge. Similarly, due to specifics of the divergent interfacial polarization, there is an inbalance in the bound charge between the polarized liquid inside and outside of the shear surface. The excess bound charge needs to be added to the excess free charge to establish the effective electrokinetic charge.  

The derivation is performed here for a spherical solute, where the geometry of the interface produces a divergent radial polarization field. The model is not directly extendable to flat interfaces studied by simulations in the past \cite{Knecht:10,Bonthuis:10}. While the polarization field is clearly inhomogeneous next to a planar surface, it often demonstrates positive and negative spikes \cite{Knecht:10,Horvath:2013fe}, which can potentially compensate each other  when the field is applied parallel to the interface to produce electrophoretic flow. The force $\langle F_x\rangle$ along the plane of the surface ($x$-axis) writes
\begin{equation}
\langle F_x\rangle =  E_x S \int \rho_b(z) dz,  
\end{equation}
where $S$ is the surface area. If the density of the bound charge $\rho_b(z)$ integrates to zero, there is no net force. In this regard, the roughness of the interface, as suggested by Knecht \text{et al} \cite{Knecht:10}, can provide the required conditions for a divergent polarization field which cannot be reduced to a one-dimensional integral shown above. 

The direct connection between the mobility of nanoparticles in water and the orientational order of the water dipoles in the interface offers opportunities for testing this prediction by experiment. One possible direction is the modification of the surface with chemical groups (surface dipoles) altering the interfacial order \cite{Giovambattista:2007kx}.  We, however, recently discovered another property dramatically affecting the  interfacial dipoles: the polarizability of the solute. Increasing the solute polarizability drives the solute-water system to the point of instability of harmonic fluctuations expressed in terms of the solvent electric field inside the solute as the order parameter. Reaching the point of global instability toward fluctuations drives a structural transition of the hydration layer, which reorients the water dipoles and creates a high density of dangling OH bonds \cite{DMprl:15}. The emergent new structure of the interface also suggests, according to Eq.\ \eqref{eq:10}, the alteration of the sign of $\sigma_0$ ($q=0$).

The results of MC simulations of HS solutes with changing isotropic dipolar polarizability $\alpha$ at the solute's center are presented in Fig.\ \ref{fig:5}. The size of the solute is maintained constant and only the polarizability is varied. One observes a switch from a positive to a negative surface charge with increasing polarizability. In other words, the isoelectric point of electrophoretic mobility can be reached, and crossed, by manipulating the polarizability of the dissolved particle. This observation opens the door to experimental testing of the model. Polarizability of semiconductor nanoparticles can be dramatically increased by photoexcitation \cite{WangHeinz:06}, which is predicted to invert the nanoparticles' mobility  (horizontal arrow in Fig.\ \ref{fig:5}b).

\textbf{Acknowledgments}.  
This research was supported by the National Science Foundation (CHE-1464810) and through XSEDE (TG-MCB080116N).

\appendix 
\section{Onsager reciprocal relations}
In order to prove the Onsager reciprocal relations for the problem of electro-osmotic current, one needs to consider the volume transport $V$ in response to the applied gradient of the external electrostatic potential $\nabla\phi_\text{ext}$ and the streaming current $I$ in response to the applied pressure gradient $\nabla p$: $V=L_{12}\nabla\phi_\text{ext}$, $I =L_{21}\nabla p$. The Onsager reciprocal relations then require $L_{12}=L_{21}$. 

We start with the equation of motion for the stationary flow of an incompressible fluid ($\nabla\cdot \mathbf{v}=0$) along the $z$-axis of a capillary \cite{Landau6}
\begin{equation}
-\eta \nabla^2 v_z + \rho (\mathbf{v}\cdot\nabla)v_z = -\nabla_z p + f_z .
\label{eqS1}  
\end{equation}
Here, $v_z(x,y)$ changes only along the cross section of the capillary ($x,y$ axes) and, therefore, $\nabla^2 = \partial^2/\partial x^2 + \partial^2/\partial y^2$. Further, $\eta$ is the viscosity and $\rho$ is the fluid density. Since no convective motion of the liquid occurs, $(\mathbf{v}\cdot\nabla)v_z$ vanishes. 

In contrast to the standard textbook description considering free charges only, the force density $f_z$ in Eq.\ \eqref{eqS1} is caused by a constant external field, $E_z = -\nabla_z \phi_\text{ext}$, applied to the entire, free and bound, charge: $f_z=\rho(\mathbf{r})E_z$, $\rho=\rho_i -\nabla\cdot \mathbf{P}$. Since the curl of $\mathbf{P}$ disappears in the divergence $\nabla\cdot\mathbf{P}$, one can put $\mathbf{P}=-\nabla\phi_b$ with the results
\begin{equation}
\rho = -\frac{1}{4\pi}\nabla^2\phi,\quad \phi=\phi_i-4\pi\phi_b,
\label{eqS2}  
\end{equation}
where $\phi_i$ is the electrostatic potential of free charges. 

We now proceed to calculate $v_z$ under the action of the force $f_z$ assuming no pressure applied to the capillary. The result from Eqs.\ \eqref{eqS1} and \eqref{eqS2} is 
\begin{equation}
v_z=-\frac{\phi_0-\phi}{4\pi\eta}E_z,
\label{eqS3}  
\end{equation}
where $\phi_0$ is the potential at the shear surface at which $v_z=0$. In standard notations $\phi_0=\epsilon_s\zeta$, where $\zeta$ is the $\zeta$-potential and $\epsilon_s$ accounts for the screening by bound charges. Here, the potential of bound charges is a part of $\phi$ and $\epsilon_s$ does not appear explicitly. A similar line of arguments can be applied to the potential of free charges $\phi_i$ connected to $\phi$ through a closure relation. When the constitutive relations of continuous dielectrics are used, one has $\phi=\epsilon_s\phi_i$, where $\phi_i$ can be determined from solving the Poisson-Boltzmann equation for the electrolyte next to the capillary wall. These details are irrelevant to our purpose since the derivation requires only the Coulomb law and the corresponding Laplace equation. 

From Eq.\ \eqref{eqS3}, one gets the volume transport
\begin{equation}
V = \int v_z dS= L_{12}\nabla_z\phi_\text{ext}
\label{eqS4}  
\end{equation}
with
\begin{equation}
L_{12} = \frac{\phi_0}{4\pi\eta}\int (1- \phi/\phi_0) dS .  
\label{eqS8}
\end{equation}

We now turn to the streaming current when the capillary is subjected to the pressure gradient $-\nabla_z p$. The current is given by the equation
\begin{equation}
I = \int v_z\rho dS = \frac{1}{4\pi}\int (\phi_0-\phi)\nabla^2 v_z dS . 
\label{eqS5}  
\end{equation}
We now put $f_z=0$ in Eq.\ \eqref{eqS1}, which results in
\begin{equation}
I= L_{21}\nabla_z p. 
\label{eqS6}  
\end{equation}
It is easy to see that
\begin{equation}
L_{21}= L_{12},
\label{eqS7}  
\end{equation}
where $L_{12}$ is given by Eq.\ \eqref{eqS8}.

\bibliographystyle{elsarticle-num} 
\bibliography{chem_abbr,dielectric,dm,statmech,protein,liquids,solvation,dynamics,simulations,surface,water,glasset,nano}

\begin{thebibliography}{10}
\expandafter\ifx\csname url\endcsname\relax
  \def\url#1{\texttt{#1}}\fi
\expandafter\ifx\csname urlprefix\endcsname\relax\def\urlprefix{URL }\fi
\expandafter\ifx\csname href\endcsname\relax
  \def\href#1#2{#2} \def\path#1{#1}\fi

\bibitem{Wall:10}
S.~Wall, The history of electrokinetic phenomena, Curr. Opin. Coll. Interf.
  Sci. 15 (2010) 119--124.

\bibitem{Hunter:81}
R.~J. Hunter, Zeta potential in colloid science, Academic Press, London, 1981.

\bibitem{Marinova:1996ud}
K.~G. Marinova, R.~G. Alargova, N.~D. Denkov, O.~D. Velev, {Charging of
  oil-water interfaces due to spontaneous adsorption of hydroxyl ions},
  Langmuir 12~(8) (1996) 2045--2051.

\bibitem{Takahashi:2005kl}
M.~Takahashi, $\zeta$-potential of microbubbles in aqueous solutions:
  Electrical properties of the gas−water interface, J. Phys. Chem. B 109~(46)
  (2005) 21858--21864.
\newblock \href {http://dx.doi.org/10.1021/jp0445270}
  {\path{doi:10.1021/jp0445270}}.

\bibitem{Beattie:09}
J.~K. Beattie, A.~M. Djerdjev, G.~G. Warr, The surface of neat water is basic,
  Farad. Disc. 141 (2009) 31--39.

\bibitem{Zimmermann:2010db}
R.~Zimmermann, U.~Freudenberg, R.~Schwei{\ss}, D.~K{\"u}ttner, C.~Werner,
  Hydroxide and hydronium ion adsorption ---a survey, Current Opinion in
  Colloid \& Interface Science 15~(3) (2010) 196--202.

\bibitem{Buch:2007kx}
V.~Buch, A.~Milet, R.~V{\'a}cha, P.~Jungwirth, J.~P. Devlin, Water surface is
  acidic, Proc Natl Acad Sci U S A 104~(18) (2007) 7342--7347.

\bibitem{Baer:2012uq}
M.~D. Baer, A.~C. Stern, Y.~Levin, D.~J. Tobias, C.~J. Mundy, Electrochemical
  surface potential due to classical point charge models drives anion
  adsorption to the air--water interface, J. Phys. Chem. Lett. 3~(11) (2012)
  1565--1570.

\bibitem{Petersen:2008ys}
P.~B. Petersen, R.~J. Saykally, Is the liquid water surface basic or acidic?
  macroscopic vs. molecular-scale investigations, Chem. Phys. Lett. 458~(4--6)
  (2008) 255--261.

\bibitem{Vacha:2011ij}
R.~V{\'a}cha, S.~W. Rick, P.~Jungwirth, A.~G.~F. de~Beer, H.~B. de~Aguiar,
  J.-S. Samson, S.~Roke, The orientation and charge of water at the hydrophobic
  oil droplet--water interface, J. Am. Chem. Soc. 133~(26) (2011) 10204--10210.

\bibitem{Yamaguchi:2015hd}
S.~Yamaguchi, T.~Tahara, {Development of Electronic Sum Frequency Generation
  Spectroscopies and Their Application to Liquid Interfaces}, J. Phys. Chem. C
  119~(27) (2015) 14815--14828.

\bibitem{Samson:2014fz}
J.-S. Samson, R.~Scheu, N.~Smolentsev, S.~W. Rick, S.~Roke, {Sum frequency
  spectroscopy of the hydrophobic nanodroplet/water interface: Absence of
  hydroxyl ion and dangling OH bond signatures}, Chem. Phys. Lett. 615 (2014)
  124--131.

\bibitem{Shapovalov:2013kc}
V.~L. Shapovalov, H.~M{\"o}hwald, O.~V. Konovalov, V.~Knecht, {Negligible water
  surface charge determined using Kelvin probe and total reflection X-ray
  fluorescence techniques}, Phys. Chem. Chem. Phys. 15~(33) (2013) 13991.

\bibitem{Joseph:08}
S.~Joseph, N.~R. Aluru, Pumping of confined water in carbon nanotubes by
  rotating-translational coupling, Phys. Rev. Lett. 101 (2008) 064502.

\bibitem{Knecht:10}
V.~Knecht, Z.~A. Levine, P.~T. Vernier, Electrophoresis of neutral oil in
  water, J. Colloid Interface Sci. 352 (2010) 223--231.

\bibitem{BenAmotz:2011cx}
D.~Ben-Amotz, {Unveiling Electron Promiscuity}, J. Phys. Chem. Lett. 2~(10)
  (2011) 1216--1222.

\bibitem{Vacha:2012dm}
R.~V{\'a}cha, O.~Marsalek, A.~P. Willard, D.~J. Bonthuis, R.~R. Netz,
  P.~Jungwirth, {Charge Transfer between Water Molecules As the Possible Origin
  of the Observed Charging at the Surface of Pure Water}, J. Phys. Chem. Lett.
  3~(1) (2012) 107--111.

\bibitem{Schoeler:2013gg}
A.~M. Schoeler, D.~N. Josephides, S.~Sajjadi, C.~D. Lorenz, P.~Mesquida,
  {Charge of water droplets in non-polar oils}, J. Appl. Phys. 114~(14) (2013)
  144903.

\bibitem{DMmp:14}
D.~V. Matyushov, Electrophoretic mobility without charge driven by polarisation
  of the nanoparticle--water interface, Mol. Phys. 112~(15) (2014) 2029---2039.

\bibitem{Bonthuis:2009kw}
D.~J. Bonthuis, D.~Horinek, L.~Bocquet, R.~R. Netz, {Electrohydraulic Power
  Conversion in Planar Nanochannels}, Phys. Rev. Lett. 103~(14) (2009) 144503.

\bibitem{Bonthuis:10}
D.~J. Bonthuis, D.~Horinek, L.~Bocquet, R.~R. Netz, Electrokinetics at aqueous
  interfaces without mobile charges, Langmuir 26~(15) (2010) 12614--12625.

\bibitem{SukAluru:10}
M.~E. Suk, N.~R. Aluru, {Suk and Aluru Reply}, Phys. Rev. Lett. 105 (2010)
  209402.

\bibitem{Daub:09}
C.~D. Daub, D.~Bratko, T.~Ali, A.~Luzar, Microscopic dynamics of the
  orientation of a hydrated nanoparticle in an electric field, Phys. Rev. Lett.
  103 (2009) 207801.

\bibitem{Tian:2014kt}
C.~S. Tian, Y.~R. Shen, {Recent progress on sum-frequency spectroscopy},
  Surface Science Reports 69~(2-3) (2014) 105--131.

\bibitem{Bonn:2015hr}
M.~Bonn, Y.~Nagata, E.~H.~G. Backus, {Molecular Structure and Dynamics of Water
  at the Water-Air Interface Studied with Surface-Specific Vibrational
  Spectroscopy}, Angew. Chem. Int. Ed. 54~(19) (2015) 5560--5576.

\bibitem{Mazur:1951vx}
P.~Mazur, J.~Overbeek, {On electro‐osmosis and streaming‐potentials in
  diaphragms: II. General quantitative relationship between electro‐kinetic
  effects}, Rec. Trav. Chim. 70 (1951) 83--91.

\bibitem{Strazdaite:2015hx}
S.~Strazdaite, J.~Versluis, H.~J. Bakker, {Water orientation at hydrophobic
  interfaces}, J. Chem. Phys. 143~(8) (2015) 084708.

\bibitem{Wen:2016df}
Y.-C. Wen, S.~Zha, X.~Liu, S.~Yang, P.~Guo, G.~Shi, H.~Fang, Y.~R. Shen,
  C.~Tian, {Unveiling Microscopic Structures of Charged Water Interfaces by
  Surface-Specific Vibrational Spectroscopy}, Phys. Rev. Lett. 116~(1) (2016)
  016101.

\bibitem{Giovambattista:2007kx}
N.~Giovambattista, P.~G. Debenedetti, P.~J. Rossky, Effect of surface polarity
  on water contact angle and interfacial hydration structure, J. Phys. Chem. B
  111~(32) (2007) 9581--9587.

\bibitem{DMjcp2:11}
D.~R. Martin, A.~D. Friesen, D.~V. Matyushov, Electric field inside a
  ``{R}ossky cavity'' in uniformly polarized water, J. Chem. Phys. 135 (2011)
  084514.

\bibitem{Nihonyanagi:2009vn}
S.~Nihonyanagi, S.~Yamaguchi, T.~Tahara, Direct evidence for orientational
  flip-flop of water molecules at charged interfaces: A heterodyne-detected
  vibrational sum frequency generation study, J. Chem. Phys. 130~(20) (2009)
  204704--5.

\bibitem{DMprl:15}
M.~Dinpajooh, D.~V. Matyushov, Interfacial structural transition in hydration
  shells of a polarizable solute, Phys. Rev. Lett. 114 (2015) 207801.

\bibitem{WangHeinz:06}
F.~Wang, J.~Shan, M.~A. Islam, I.~P. Herman, M.~Bonn, T.~F. Heinz, Exciton
  polarizability in semiconductor nanocrystals, Nat. Mater. 5~(11) (2006)
  861--864.

\bibitem{Landau8}
L.~D. Landau, E.~M. Lifshitz, Electrodynamics of {C}ontinuous {M}edia,
  Pergamon, Oxford, 1984.

\bibitem{Gubbins:84}
C.~G. Gray, K.~E. Gubbins, Theory of Molecular Liquids, Clarendon Press,
  Oxford, 1984.

\bibitem{Jackson:99}
J.~D. Jackson, Classical {E}lectrodynamics, Wiley, New York, 1999.

\bibitem{EygesBook:72}
L.~Eyges, The Classical Electromagnetic Field, Dover Publications, Ney York,
  1972.

\bibitem{Ballenegger:05}
V.~Ballenegger, J.-P. Hansen, Dielectric permittivity profiles of confined
  polar liquids, J. Chem. Phys. 122 (2005) 114711.

\bibitem{Horvath:2013fe}
L.~Horv{\'a}th, T.~Beu, M.~Manghi, J.~Palmeri, {The vapor-liquid interface
  potential of (multi)polar fluids and its influence on ion solvation}, J.
  Chem. Phys. 138~(15) (2013) 154702.

\bibitem{Bonthuis:2013fk}
D.~J. Bonthuis, R.~R. Netz, Beyond the continuum: How molecular solvent
  structure affects electrostatics and hydrodynamics at solid-electrolyte
  interfaces, J. Phys. Chem. B 117~(39) (2013) 11397--11413.
\newblock \href {http://dx.doi.org/10.1021/jp402482q}
  {\path{doi:10.1021/jp402482q}}.

\bibitem{Ohshima:06}
H.~Ohshima, Theory of Colloid And Interfacial Electric Phenomena, Academic
  Press, London, 2006.

\bibitem{DMcpl:11}
A.~D. Friesen, D.~V. Matyushov, Local polarity excess at the interface of water
  with a nonpolar solute, Chem. Phys. Lett. 511 (2011) 256--261.

\bibitem{Remsing:2014fo}
R.~C. Remsing, M.~D. Baer, G.~K. Schenter, C.~J. Mundy, J.~D. Weeks, {The Role
  of Broken Symmetry in Solvation of a Spherical Cavity in Classical and
  Quantum Water Models}, J. Phys. Chem. Lett. 5 (2014) 2767--2774.

\bibitem{Beck:2013gp}
T.~L. Beck, {The influence of water interfacial potentials on ion hydration in
  bulk water and near interfaces}, Chem.\ Phys.\ Lett. 561-562 (2013) 1--13.

\bibitem{Luzar:1987gf}
A.~Luzar, D.~Bratko, L.~Blum, {Monte Carlo simulation of hydrophobic
  interaction}, J. Chem. Phys. 86~(5) (1987) 2955--2959.

\bibitem{Scatena:2001ve}
L.~F. Scatena, G.~L. Richmond, Orientation, hydrogen bonding, and penetration
  of water at the organic/water interface, J. Phys. Chem. B 105~(45) (2001)
  11240--11250.

\bibitem{Strazdaite:2014jc}
S.~Strazdaite, J.~Versluis, E.~H.~G. Backus, H.~J. Bakker, {Enhanced ordering
  of water at hydrophobic surfaces}, J. Chem. Phys. 140~(5) (2014) 054711.

\bibitem{com:Bonthuis}
No mobility at zero charge was found in Ref.\ 21 considering a planar interface
  with a constant external field applied along the plane. The polarization of
  the interface was modeled with an inhomogeneous field $E_z(z)$ applied
  normally to the plane and causing the polarization $P_z = \alpha E_z$.
  According to our arguments, this setup creates the bound charge $\rho_b =
  -\alpha\partial E_z/\partial z$, which is proportional to the density of the
  external charge per Maxwell equation. Since no external charges are placed in
  water, $\rho_b = 0$ and there is no net force, in agreement with their
  conclusions.

\bibitem{Bonthuis:11}
D.~J. Bonthuis, K.~F. Rinne, K.~Falk, C.~N. Kaplan, D.~Horinek, A.~N. Berker,
  L.~Bocquet, R.~R. Netz, Theory and simulations of water flow through carbon
  nanotubes: prospects and pitfalls, J. Phys.: Condens. Matter 23 (2011)
  184110.

\bibitem{Einstein05}
A.~Einstein, Investigations on the theory of the Brownian movement, BN
  Publishing, 2011.

\bibitem{Bratko:2007jl}
D.~Bratko, C.~D. Daub, K.~Leung, A.~Luzar, Effect of field direction on
  electrowetting in a nanopore, J. Am. Chem. Soc. 129~(9) (2007) 2504--2510.
\newblock \href {http://dx.doi.org/10.1021/ja0659370}
  {\path{doi:10.1021/ja0659370}}.

\bibitem{Kihara:58}
T.~Kihara, Intermolecular forces and equation of state of gases, Adv. Chem.
  Phys. 1 (1958) 267--307.

\bibitem{DMjcp2:15}
M.~Dinpajooh, D.~V. Matyushov, Free energy of ion hydration: Interface
  susceptibility and scaling with the ion size, J. Chem. Phys. 143 (2015)
  044511.

\bibitem{tip3p:83}
W.~L. Jorgensen, J.~Chandrasekhar, J.~D. Madura, R.~W. Impey, M.~L. Klein,
  Comparison of simple potential functions for simulating liquid water, J.
  Chem. Phys. 79~(2) (1983) 926--935.

\bibitem{Sarupria:09}
S.~Sarupria, S.~Garde, Quantifying water density fluctuations and
  compressibility of hydration shells of hydrophobic solutes and proteins,
  Phys. Rev. Lett. 103 (2009) 037803.

\bibitem{Falkenhagen}
H.~Falkenhagen, Electrolytes, Clarendon Press, Oxford, 1934.

\bibitem{Knecht:2013jo}
V.~Knecht, B.~Klasczyk, R.~Dimova, {Macro- versus Microscopic View on the
  Electrokinetics of a Water--Membrane Interface}, Langmuir 29~(25) (2013)
  7939--7948.

\bibitem{Pieniazek:2011fc}
P.~A. Pieniazek, C.~J. Tainter, J.~L. Skinner, {Interpretation of the water
  surface vibrational sum-frequency spectrum}, J. Chem. Phys. 135~(4) (2011)
  044701.

\bibitem{Roy:2014kl}
S.~Roy, S.~M. Gruenbaum, J.~L. Skinner, {Theoretical vibrational sum-frequency
  generation spectroscopy of water near lipid and surfactant monolayer
  interfaces}, J. Chem. Phys. 141~(18) (2014) 18C502.

\bibitem{Lynden-Bell:1997uq}
R.~M. Lynden-Bell, J.~C. Rasaiah, From hydrophobic to hydrophilic behaviour: A
  simulation study of solvation entropy and free energy of simple solutes, J.
  Chem. Phys. 107~(6) (1997) 1981--1991.

\bibitem{Drzymala:1999bs}
J.~Drzymala, Z.~Sadowski, L.~Holysz, E.~Chibowski, Ice/water interface: Zeta
  potential, point of zero charge, and hydrophobicity, J. Colloid Interface
  Sci. 220~(2) (1999) 229--234.

\bibitem{Landau6}
L.~D. Landau, E.~M. Lifshits, {Fluid Mechanics}, 2nd Edition, Elsevier,
  Amsterdam, 2004.

\end{thebibliography}

\end{document}